\documentstyle[prl,aps,epsfig,twocolumn,subfigure,graphicx]{revtex} 
\begin{document}
\title{ Energy Dependence of Nuclear Transparency in C(p,2p) Scattering} 
 
\draft
\author{
A. Leksanov$^e$,
J. Alster$^a$, G. Asryan$^{c,b}$,
Y. Averichev$^{h}$,
D. Barton$^c$, V. Baturin$^{e,d}$, N. Bukhtoyarova$^{c,d}$,
A. Carroll$^c$,
S. Heppelmann$^e$,
T. Kawabata$^f$,
Y. Makdisi$^c$, 
A. Malki$^a$,
E. Minina$^e$, I. Navon$^a$,
H. Nicholson$^g$, A. Ogawa$^e$,
Yu. Panebratsev$^{h}$,
E. Piasetzky$^a$,
A. Schetkovsky$^{e,d}$,
 S. Shimanskiy$^{h}$,
A. Tang$^i$, J.W. Watson$^i$,
H. Yoshida$^f$,
 D. Zhalov$^e$
}
\address{
 $^a$School of Physics and Astronomy, Sackler Faculty of Exact
  Sciences, Tel Aviv University, Ramat Aviv 69978, Israel,
 $^b$Yerevan Physics Institute, Yerevan 375036, Armenia,
 $^c$Collider-Accelerator Department, Brookhaven National 
  Laboratory,Upton, NY 11973, USA,
 $^d$Petersburg Nuclear Physics Institute, Gatchina, St. Petersburg 188350, Russia,
 $^e$Physics Department, Pennsylvania State University,University Park,
 PA 16801, USA,
 $^f$Department of Physics, Kyoto University, Sakyoku, Kyoto, 606-8502, Japan,
 $^g$Department of Physics, Mount Holyoke College, South Hadley, MA 01075, USA, 
 $^h$J.I.N.R., Dubna, Moscow 141980, Russia,
 $^i$Department of Physics, Kent State University, Kent, OH 44242, USA
}
\date{\today}
\maketitle
\begin{abstract}
The transparency of carbon for (p,2p) quasi-elastic 
events was measured at beam momenta ranging 
from 5.9 to 14.5 GeV/c at 90$^\circ$ c.m. 
The four-momentum transfer squared ( $Q^2$)
ranged from 4.7 to 12.7 (GeV/c)$^2$. 
We present the observed beam momentum dependence of the 
ratio of the carbon to hydrogen cross sections. We also
apply a model for the nuclear momentum distribution
of carbon to obtain the nuclear transparency.
We find a sharp rise in 
transparency as the beam momentum is increased to 9 GeV/c and a reduction to
approximately the Glauber level at higher energies. 
\end{abstract}
\narrowtext

This paper reports a new measurement of the transparency of 
the carbon nucleus in the C(p,2p) quasi-elastic scattering process 
near 90$^\circ$ $pp$ center of mass (c.m.). 
These new data verify a surprising beam momentum dependence 
that was first observed most clearly 
with aluminum targets in 1988 \cite{Carroll:1988rp}.
While the original result was very provocative, that measurement involved 
momentum analysis of only one of the two final-state protons,
raising some questions about the quality of event selection.
We now report on a new measurement of carbon quasi-elastic scattering 
with the EVA detector\cite{kn:lev1} at the Brookhaven AGS. 
This       
cylindrically-symmetric large-angle
tracking spectrometer, with a 3 m long super-conducting solenoid magnet, provided
symmetrical momentum and angle reconstruction of the two final-state protons.
Initial results with this apparatus\cite{Mardor:prl98} 
emphasized the angular dependence of the transparency at 
the lower beam momenta of
5.9 and 7.5 GeV/c.
We now present a newer measurement of the energy dependence of transparency 
for beam momenta ranging from 5.9 to 14.5 GeV/c.

Color transparency refers to a QCD phenomenon, predicted in 1982
by Brodsky \cite{brod:82} and Mueller \cite{mue:82}, involving 
reduction of secondary absorption in proton-nucleus quasi-elastic scattering. 
These theorists deduced that when a proton traversing 
the nucleus experiences a hard collision, a special quantum state is selected.
That special state involves the part of the proton wave function 
that is most ``shock-resistant`` and that tends to survive the 
hard collision without breaking up or radiating a gluon. 
This state is also expected to have a reduced
interaction with the spectators in the target nucleus. 
The state is predicted to involve a rare component of the proton 
wave function that is dominated by three valence quarks at small 
transverse spatial separation.
The color transparency prediction is that the 
fraction of nuclear protons contributing to (p,2p) quasi-elastic scattering
should increase from a level consistent with Glauber absorption
\cite{Glauber:59,Frankfurt:1994nn}
at low $Q^2$ to near unity at higher $Q^2$. 
 
The fundamental sub-process in the quasi-elastic events is a $pp$ 
interaction.
The quasi-elastic events are characterized by
a small missing energy ($E_{F}$) and momentum ($\vec{P}_{F}$), defined in terms
of the initial and final-state energies and momenta $E_i$ and
$\vec{P}_i$ (\textit{i}=1,2 for  beam and target protons and  \textit{i}=3,4 for final-state protons)
\begin{eqnarray}&
E_F =E_3+E_4-E_1 - m_p\label{pmiss}&\\
&\vec{P}_F=\vec{P}_3+\vec{P}_4-\vec{P}_1,~~~~m_{M}^2=E_F^2-\vec{P}_F^2.&\nonumber
\end{eqnarray}

In the spirit of the impulse approximation,
we  identify the missing momentum of Equation \ref{pmiss}
with the momentum of the nucleon in the nucleus while 
recognizing that in the transverse direction this relation is 
spoiled by elastic re-scattering.
Because the $90^\circ$ c.m. $pp$ cross section
strongly depends on one longitudinal
light-cone
component 
of the missing momentum,
we express the missing momentum in light-cone coordinates
with the transformation 
$(E_F,P_{Fz}) \rightarrow (E_F + P_{Fz},E_F - P_{Fz})$. The coordinate system
takes $\hat{z}$ as the beam direction and $\hat{y}$ 
normal to  the scattering plane. The four-dimensional  volume element is
\begin{equation}
dE_F~d^3 \vec{P}_F \rightarrow d^2\vec{P}_{FT} \frac{d\alpha}{\alpha} d (m_{M}^2)\label{jacob}\\
\label{arreq}
\end{equation}
where $\vec{P}_{FT}$ is the transverse part of the missing momentum vector.
The ratio $\frac{\alpha} {A}$ 
is associated with the fraction of light-cone momentum 
carried by a single proton in a nucleus with $A$ nucleons,
\begin{equation}
\alpha \equiv A \frac{(E_F-P_{Fz})}{M_A} \simeq 1-\frac{P_{Fz}}{m_p}.
\label{one} 
\end{equation}

Elastic $pp$ scattering occurs at a singular point
($m_{M}^2=0$, $P_{FT}^2=0$, $\alpha=1$) in this four-dimensional
phase-space while the quasi-elastic process produces a broader distribution
about the same point.
The kinematic cuts used to define event candidates are  
summarized as follows:
\begin{eqnarray}
 |P_{Fx}|< 0.5 \frac{GeV}{c}; {~} |P_{Fy}|< 0.3 \frac{GeV}{c}; ~|1-\alpha_0|<0.05 \label{cuts}\\
\alpha_0 \equiv 1-\frac{\left(\sqrt{(E_1+m_p)^2-4 m_p^2}\right)
                       \cos{(\frac{\theta_3+\theta_4}{2})}-P_1}{m_p}\nonumber.
\end{eqnarray}
Taking into account the measurement resolution,
our best determination of the light-cone momentum
in the kinematic region of interest is obtained
by measuring $\alpha_0$ instead of $\alpha$ directly.
The variable $\alpha_0$ is an approximation to $\alpha$ that, 
for fixed beam energy, depends only on final-state lab polar angles 
$\theta_3$ and $\theta_4$.
Simulations indicate that in the kinematic region of interest
near $\alpha=1$ and near 90$^\circ$ c.m.,
the difference between
$\alpha_0$ and $\alpha$ is less than 0.005. 
In the following analysis, the
 experimental error in the measurement of $\alpha$ using
the $\alpha_0$ variable
is about 1.5\%. 
This is the same set of cuts used in 
previously published analysis \cite{Mardor:prl98} where
the emphasis was on the c.m.  angle dependence of transparency. 
Here the transparency is analyzed at 5 beam
momenta,  $(5.9,8.0,9.1,11.6,14.4)$ GeV/c, and the
c.m. angular range for each beam momentum is from $\theta_{low}$ to  
$90^\circ$ where  $\theta_{low}$ is 
$(86.2^\circ,87.0^\circ,86.8^\circ,85.8^\circ, 86.3^\circ)$
at each corresponding momentum.

The elastic or quasi-elastic  event selection procedure 
involves first the application
of the cuts of Eq. \ref{cuts}, associated with three of the four
missing energy-momentum relations. 
In the previous 5.9 and 7.5 GeV/c analysis, the signal/background separation
was extracted from 
the missing-energy distribution. 
A model for the background distribution, based on observed
events with additional soft-track production in the detector provided 
guidance for the shape of the background distributions.
The use
of the missing-energy distribution
for extraction of signal from background is
less satisfactory for this analysis. 
The missing-energy resolution
varies with beam momentum, degrading 
from about 300 MeV to 500-700 MeV as beam
momentum increases.
Furthermore, the phase-space available for inclusive-event
production falls rapidly to zero as the missing energy
approaches zero.
Thus, most of the background
is under the resolution-dependant tail 
of the quasi-elastic signal.

We now describe an improved
analysis procedure where the background subtraction
utilizes the variation in the density of measured events per unit
four-dimensional missing-momentum space, a distribution which
shows a sharp 
quasi-elastic peak at missing four-momentum of zero with a very 
smooth background.
Noting that because we are cutting tightly only on 
$\alpha$, we can observe the peaking signal 
over background in the remaining three dimensions
of momentum space.
From the form of the missing four-momentum 
differential element shown in Eq.
\ref{jacob},
we note that for any selected region of  $\alpha$, 
the selected  four-momentum volume is proportional
to $\Delta P_{FT}^2 \times \Delta m_M^2$. In a 2D distribution of 
$m_M^2$ vs $P_{FT}^2$, each equal area corresponds to an equal volume
of this momentum space. We introduce the variable
$P^4 \equiv m_{M}^{4} +P_{FT}^{4}$, the square of the radial distance from
the origin 
in the  $P_{FT}^2 \times m_M^2$ plane.
Each equal interval in
$\Delta P^4$ also
corresponds to an equal volume of missing four-momentum.
The motivation for replacing the missing-energy distribution
with the $P^4$ distribution for signal background extraction
is the expectation that inclusive background may be a
smoother and flatter distribution and the signal will
be sharper.

\begin{figure}[h!]
\centering
\includegraphics[width=3.5 in]{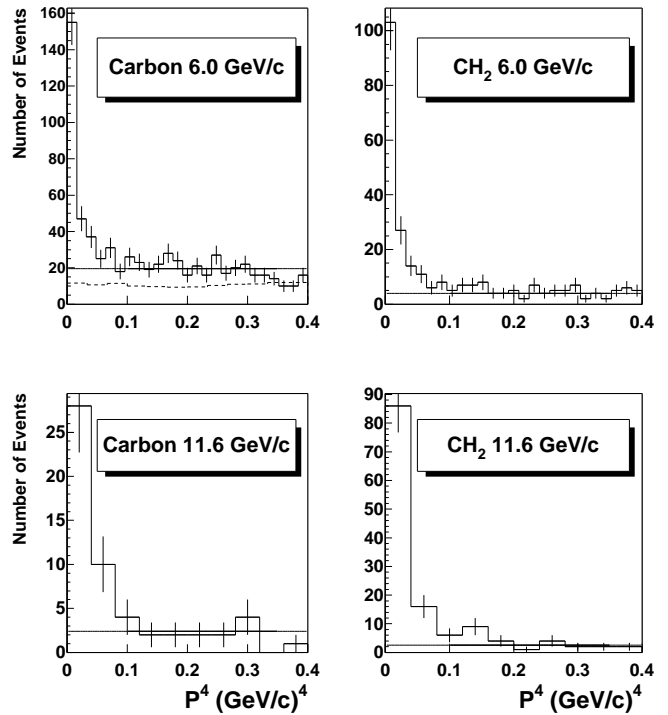}
\caption{ The $P^{4}$ distributions
of  carbon  and CH$_{2}$ events that satisfy the cuts defined by 
Eq. \ref{cuts}. Distributions
at beam momenta of 5.9 and 11.6 GeV/c are shown. The solid line
indicates the constant background level from fits to the off-peak region. 
In the upper-left frame, the dashed line indicates the distribution obtained 
when the data selection cuts are applied to the tagged background events
discussed in the text.}
\label{fig:p4}
\end{figure}

In Figure \ref{fig:p4} we show the histogram of $P^4$ for
the data sets taken at 5.9 and 11.6 GeV/c  for both  
carbon and CH$_{2}$. These events were selected to have exactly
two charged tracks and to pass 
the cuts described in Eq. \ref{cuts}.
To verify that the background $P^4$ distribution is smooth near $P^4=0$, 
we also study a class of tagged inclusive events that satisfy the same
selection
cuts but also produce soft charged tracks in 
the spectrometer inner chambers.
The tagged inclusive distribution for 5.9 GeV/c carbon data 
is plotted with a dashed line.
For these tagged background events, the distribution in $P^4$ is
constant to within about 10\%.
The number of such tagged background events observed at 11.6 GeV/c
is too small to analyze but the few events seen are again
consistant with a flat distribution.
The distribution of tagged background events represents 
our best determination of the distribution of the 
inclusive background under the quasi-elastic peak, for which
no extra charged tracks are observed.
We can conclude that this selection
process, including the cuts of Eq. \ref{cuts}, does not induce an 
enhancement in the background near $P^4=0$.

For extraction of transparency, a
constant background level is fit to the
distribution in the region $0.15<P^4<0.35$. The background under
the peak in the  $0<P^4<.1$ region ranges from 15\% to 25\% of the signal
at different
beam energies. We estimate the systematic error in the determination
of background to be about 25\% resulting in systematic errors in
the extracted signals of
about 5\%.
This compares favorably 
with the 1988 analysis where the background was typically
greater than 100\% of the signal.
We also note the there is no systematic
difference in the  transparency obtained
from this analysis of the  $P^4$ distribution 
as compared to the analysis of the 
missing-energy distribution used
in previous publications. However, the background
for the missing-energy analysis
is a larger fraction of the signal and the background shape
is poorly determined for data at
higher beam momentum.

\begin{figure}[h!]
\centering
{
\epsfig{figure=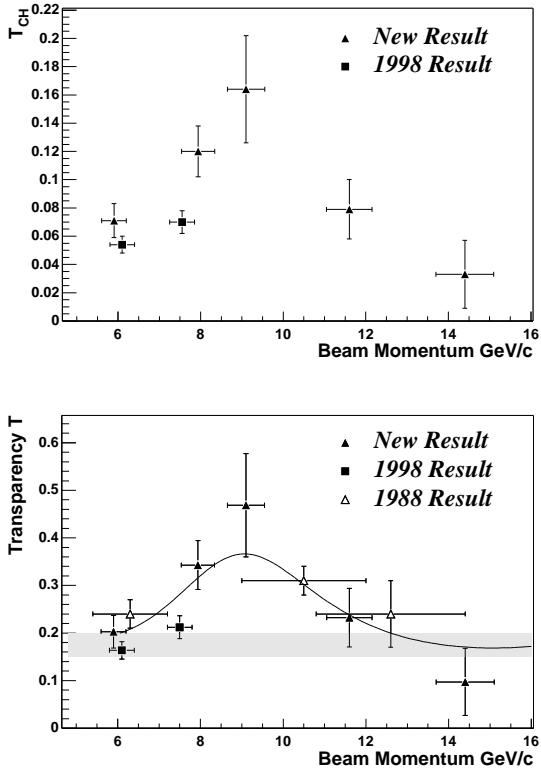, width=8.cm}
}
\vspace{.1cm}
\caption{ 
TOP: The transparency ratio $T_{CH}$ as a function of the beam
 momentum for both the 
present result and two points from the 1998 publication [3].
BOTTOM: The transparency $T$ versus beam momentum.
The vertical errors shown here are all statistical errors, 
which dominate for these measurements. The horizontal errors reflect
the $\alpha$ bin used.
The shaded band represents the Glauber calculation for carbon [9].
The solid curve shows
the shape $R^{-1}$ as defined in the text.
The 1998 data cover the c.m. angular region
from $86^\circ-90^\circ$. 
For the new data, a similar angular region is covered as
is discussed in the text. 
The 1988 data cover $81^\circ-90^\circ$ c. m.
}
\label{fig2}
\end{figure}

We define $T_{CH}$ to be the experimentally-observed ratio of the carbon 
event rate to the hydrogen event rate per target proton for events 
satisfying the 
specific set of kinematic 
cuts given in Eq.\ \ref{cuts}. 
The normalization of this
ratio depends upon the cuts used and upon the nuclear momentum
distribution. However, with the restriction to the region
near $\alpha=1$, 
the energy dependence of $T_{CH}$ closely tracks the 
energy dependence of the actual transparency $T$. 
The wide range of accepted 
transverse momentum insures that non-absorptive 
secondary interactions are
included in the event selection.
We determine $R_C$ and $R_{CH_2}$, the elastic or quasi-elastic event rates
per beam proton and per carbon atom, for sets of data taken at each 
beam momentum on $CH_2$ and carbon targets.
The experimental ratio, $T_{CH}$ is
\begin{equation}
T_{CH} = \frac{1}{3} \frac{R_C}{R_{CH_{2}}-R_C}. 
\end{equation}
The values of
$T_{CH}$ which are plotted in Fig. \ref{fig2}(top) show a significant
beam momentum dependence.

To extract the transparency $T$,
we will also introduce a relativistic nuclear 
momentum distribution function that specifies
the differential probability density per unit
four-momentum to observe a particular missing energy
and momentum.
Implicitly integrating over the missing mass 
$(m_{M})^2$, we characterize the nuclear momentum distribution over 
light-cone fraction and transverse momentum, $n(\alpha,\vec{P}_{FT})$. 
We also introduce the integral of this distribution function over
the transverse coordinates:
\begin{equation}
N(\alpha)=\int \int d\vec{P}_{FT} n(\alpha,\vec{P}_{FT}).
\label{Nalpha}
\end{equation}
The distribution functions $N(\alpha)$ can be estimated from non-relativistic
nuclear momentum distributions.  We will refer to $n_C(P)$, a recent parameterization
of a spherically-symmetric carbon nuclear momentum distribution 
by Ciofi degli Atti $et$ $al.$\cite{CiofidegliAtti:1996qe}. 

The nuclear transparency $T$ measures the reduction in the
quasi-elastic scattering cross section in comparison to the elastic
cross section due to
initial and final-state interactions with the spectator nucleons.
It can be defined in terms of
the experimentally-observed 
ratio $T_{CH}$ through a convolution of the fundamental $pp$ cross section with
a nuclear distribution function $n(\alpha,\vec{P}_{FT})$ and the $pp$ elastic
cross section  $\frac{d 
\sigma}{dt}_{pp}(s)$.
In terms of $s$ and $s_0$ defined below,
\begin{equation}
T_{CH}=T 
\int_{}^{} d\alpha \int d^2 \vec{P}_{FT}~
n(\alpha,\vec{P}_{FT}) \frac{
\frac{d \sigma}{dt}_{pp} \left( s( \alpha ) \right) }
{\frac{d \sigma}{dt}_{pp} \left( s_0 \right) }
\label{six}
\end{equation}
where the c.m. energy squared for elastic and quasi-elastic
scattering is $s_0  =  2 m _p E_1+2 m_p^2$ and 
$s(\alpha)  \simeq  \alpha s_0$.



%
%
Because distributions in $\vec{P}_{FT}$ and $\alpha$ will
be weighted by the $pp$ cross section, 
the distribution is skewed toward small $\alpha$. 
In the kinematic region of interest, the c.m.  energy of the 
$pp \rightarrow pp$ sub-process will be nearly independent of 
$\vec{P}_{FT}$ but will depend critically upon $\alpha$.

The energy dependence of 
$T_{CH}$ (Fig.\ \ref{fig2}(top)) 
and $T$ (Fig. \ref{fig2}(bottom)) are both presented here.
We emphasize that the
striking energy dependence of transparency
is seen in the simple ratio of event rates without
assumptions about the nuclear momentum distribution.
Fig. \ref{fig2}(bottom) also shows the comparison to the 
carbon measurement that was
reported in
our 1988 paper.  The 1988 data have been re-normalized to use 
the nuclear momentum distributions of Ref. \cite{CiofidegliAtti:1996qe}.
The comparison demonstrates the consistent pattern for
a peaking of the transparency at beam momentum of 9 to 10 GeV/c, and
a return to Glauber levels at 12 GeV/c and above.
The Glauber prediction and uncertainty associated with it, as calculated \cite{Frankfurt:2001ty}, is shown as a shaded band in Fig. \ref{fig2}(bottom). The
probability that our new result with carbon is consistent with the band of
Glauber values is less than 0.3$\%$, and compared to a best constant fit
of 0.24 the probability is less than 0.8$\%$. 

Several modifications of the original prediction for energy dependence of 
color transparency have been discussed \cite{Frankfurt:1994nn,kopeliovich:prl70,miller:prd44}. 
One model directly applicable to this measurement has been
suggested by Ralston and Pire\cite{Ralston:1988rb}.
They noted that the short-distance contribution to the 90$^\circ$(c.m.)
cross section is predicted to have a $s$ dependence of $s^{-10}$.  
Other softer contributions to the cross section 
result in deviations from scaling by as much as a factor of two. They
predict that the interference between these processes produces an
oscillatory cross section and  transparency.  
Parameterizing  $R(s)$, the ratio of observed $pp$ cross section to
the $s^{-10}$ scaling prediction, with their model,
Ralston and Pire argue that the energy dependence of transparency should 
reflect the shape of $R^{-1}(s)$.
We have included the curve $R^{-1}(s)$ as the solid line 
on Fig. \ref{fig2}(bottom) with arbitrary normalization.

Another perspective on the $s$ dependence was suggested by Brodsky and 
de Teramond \cite{Brodsky:1988xw}.  They suggest that the energy dependent 
structure in $R(s)$, with excess cross section
above 10 GeV/c and the corresponding reduction in the transparency,
could be related to a resonance or threshold for a new scale of physics.
They point out that the  open-charm threshold 
is in this region.
A measurement of transparency with polarized  beams and targets 
should distinguish between
these models \cite{Crosbie:1981tb}.

\vspace{-.5cm}
Nuclear transparency has been measured with electron beams at SLAC\cite{O'Neill:1995mg} at $Q^2$ up to 6.8 GeV$^2$ corresponding to about 8 GeV/c of
beam momentum in this (p,2p) measurement.
No clear disagreement with the Glauber model was seen
in (e,e'p) measurement.
In has been argued, however,
that in this $Q^2$ region
the apparent disagreement\cite{Frankfurt:1994nn,Frankfurt:1994hf} 
can be explained within a unified model of the 
time evolution of the interacting proton state.
The authors claim that for some choices of  model parameters,
higher $Q^2$ is  required for observations with electrons.

In conclusion, we confirm the striking energy dependence
observed in the 1988 measurement.
We have extended the measurement of transparency to higher energy and have
shown that the anomalous beam momentum dependence originally observed
most clearly in aluminum is similar for carbon targets.
While the peaking of transparency in the 8 to 9 GeV/c region 
corresponds to about twice the Glauber levels, 
the return to Glauber in the 
12 to 15 GeV/c region is established.

This research was supported by the U.S. - Israel Binational Science
Foundation, the Israel Science Foundation founded by the Israel Academy of
Sciences and Humanities, NSF grants PHY-9804015, PHY-0072240 and
the U.S. Department
of Energy grant DEFG0290ER40553.

\end{document}